\documentclass{INTERSPEECH2023}
\newcommand{\rev}[1]{{\color{black} #1}}
\interspeechcameraready

\title{Parameter Selection for Analyzing Conversations with Autism Spectrum Disorder}
\name{Tahiya Chowdhury$^1$, Veronica Romero$^{1,2}$, Amanda Stent$^1$}
\address{
  $^1$Davis Institute for AI, Colby College, USA\\
  $^2$Department of Psychology, Colby College, USA}
\email{tahiya.chowdhury@colby.edu, veronica.romero@colby.edu, ajstent@colby.edu}

\begin{document}

\maketitle
 
\begin{abstract}
% 1000 characters. ASCII characters only. No citations.
%Manuscripts submitted to INTERSPEECH 2023 must use this document as both an instruction set and as a template. Do not use a past paper as a template. Always start from a fresh copy, and read it all before replacing the content with your own.

%Before submitting, check that your manuscript conforms to this template. If it does not, it may be rejected. Do not be tempted to adjust the format! Instead, edit your content to fit the allowed space. The maximum number of manuscript pages is 5. The 5th page is reserved exclusively for references, which may begin on an earlier page if there is space.

%The abstract is limited to 1000 characters. The one in your manuscript and the one entered in the submission form must be identical. Avoid non-ASCII characters, symbols, maths, italics, etc as they may not display correctly in the abstract book. Do not use citations in the abstract: the abstract booklet will not include a bibliography.  Index terms appear immediately below the abstract. 
The diagnosis of autism spectrum disorder (ASD) is a complex, challenging task as it depends on the analysis of interactional behaviors by psychologists rather than the use of biochemical diagnostics. In this paper, we present a modeling approach to ASD diagnosis by analyzing acoustic/prosodic and linguistic features extracted from diagnostic conversations between a psychologist and children who either are typically developing (TD) or have ASD. We compare the contributions of different features across a range of conversation tasks. We focus on finding a minimal set of parameters that characterize conversational behaviors of children with ASD. Because ASD is diagnosed through conversational interaction, in addition to analyzing the behavior of the children, we also investigate whether the psychologist's conversational behaviors vary across diagnostic groups. Our results can facilitate fine-grained analysis of conversation data for children with ASD to support diagnosis and intervention.

\end{abstract}


\section{Introduction}
%\section{Related Work}

Autism spectrum disorder (ASD) refers to a range of developmental disabilities characterized by deficits in social communication and interaction. In the absence of therapeutic intervention, adolescents with ASD exhibit impairments in social interaction throughout their lives~\cite{american_psychiatric_organization_diagnostic_2013}, including an inability to display or reciprocate verbal aspects of communication appropriately and, in turn, successfully maintain effective social interactions~\cite{tager-flusberg_language_2007, tager-flusberg_psychological_1999, mundy_nature_1997, landa_social_2000}. The diagnosis of ASD is a complex, challenging task as it depends on behavioral symptoms identified by psychologists with no reliable biochemical diagnostic tests available. Existing diagnostic instruments, such as the Autism Diagnostic Observation Schedule (ADOS)~\cite{lord_autism_2000}, rely on qualitative coding by expert assessors for the presence or absence of certain behavioral markers across multiple structured conversational scenarios. The assessor has to simultaneously engage the child in conversation, monitor their own conversational behavior, and make diagnostic notes. AI-based techniques have the potential to reduce the cognitive load on the assessor by providing objective measurements of conversational behaviors and thus to augment the assessor's workflow.

Previous research has explored the efficacy of ML tools using speech and language features for identifying atypical and behavioral signals of ASD in communication from conversation data~\cite{fusaroli_hearing_2019, fusaroli2017voice, fusaroli2022toward}. Acoustic-prosodic features such as pitch~\cite{kiss2012quantitative}, intonation, and rhythm~\cite{bone2015acoustic} derived from diagnostic conversations %such as conversations about feelings or story-telling narratives%
have been found useful for identifying ASD individuals. Language features such as word usage~\cite{song2021natural}, social and cognitive linguistic word counts~\cite{kumar16b_interspeech}, and semantic similarity~\cite{goodkind2018detecting} have also been investigated and found to contain behavioral markers helpful to distinguish between ASD and TD children. While these findings are promising, prior work uses conversation samples from a limited set of diagnostic tasks and may not generalize to different conversational contexts. 

Prior research has also explored the use of information from the conversational partner's interactions to predict diagnosis outcome. The conversational partner's acoustic-prosodic cues have been found to be predictive of ASD symptom severity~\cite{bone2014psychologist} and engagement levels~\cite{10.1016/j.csl.2015.09.003}. \cite{kumar16b_interspeech} found a significant correlation between the conversational partner's language use and ASD severity score. \rev{~\cite{muskett2010inflexibility} used social context in conversations to explore its relationship with symptoms of autism.} However, these findings also come from a limited set of diagnostic tasks.

In this work, we aim to answer the following questions:

\begin{itemize}
    \item What are the most informative acoustic-prosodic indicators of children with ASD in different conversational contexts?
    \item What are the most informative linguistic indicators of children with ASD in different conversational contexts?
    \item Can we classify ASD and TD children using a minimal set of characteristics of the child's spoken language?
    \item Can we classify ASD and TD children using a minimal set of  characteristics of the interlocutor's spoken language?
\end{itemize}

\section{Data}
We used data collected during sessions of the Autism Diagnostic Observation Schedule - Second Edition (ADOS-2), an assessment tool used to categorize ASD impairment~\cite{lord_autism_2000}. In this assessment, a child and a certified adult assessor (usually a psychologist) engage in a sequence of semi-structured activities to assess behavioral patterns of the child. Our data includes conversations spanning 14 different subtasks from Module 3 of the ADOS-2, which is designed for verbally fluent children and adolescents (Table~\ref{tab:demography}). Depending on the subtask, the child may be asked to tell a story, play with toys, act out a cartoon, or discuss topics such as emotion, loneliness, and friends. Because we are interested in different conversational contexts, we included all 14 tasks from the module, which are listed in Table~\ref{tab:classification_child_exp_best}.

Our data included 29 children (14 ASD, 15 TD) whose age ranged between 10-15 years. Each ADOS-2 assessment session lasted on average 40-60 minutes; all sessions were administered by the same psychologist. For this work, we first prepared a separate recording of each subtask using annotations done by a research assistant from the video recordings of the sessions. The average length of these videos is 5 minutes. We used pyannote~\cite{Bredin2020} to perform speaker diarization. We then used PyDub~\cite{pydub} to partition the subtask audio into shorter segments based on speaker identity and turn boundaries derived using pyannote. This allowed us to automatically separate the child and psychologist turns into separate audio recordings, resulting in a total of 6440 utterances. % AS - what does NAN mean in this context? empty transcript?
After removing the empty utterances, our final set contained 116 utterances per child on average. %and 113 on average for experimenter%
We then used the Whisper open-source speech recognizer~\cite{radford2022robust} to transcribe each turn.

\begin{table}[t!]
  \caption{Demographic information of the ADOS-2 dataset.}
  \label{tab:demography}
  \centering
  \begin{tabular}{ r@{}l  r }
    \toprule
    \multicolumn{2}{c}{\textbf{Category}} & 
                                         \multicolumn{1}{c}{\textbf{Statistics}} \\
    \midrule
    Age (years)                     & & Range: 10-15 (mean: 12.27, std: 1.75)~~~\\
    Gender                       & & 21 male, 8 female~~~             \\
    Race                       &  & White: 20, Afr. Am.: 6, Hisp.: 2, Asian: 1~~~      \\
    Diagnosis         & & ASD: 14, TD: 15~~~\\
    \bottomrule
  \end{tabular}
  
\end{table}

%%%%%% corrected result after Bonferroni Correction%%%%%%%%
\begin{table*}[thp]
  \caption{Correlation coefficients of statistically significant indicators of ASD in ADOS-2 tasks as assessed using the conversation turns of child conversation participants. \rev{We include only the tasks for which at least one feature was found significant after Holm's sequential Bonferroni correction.}}
  \label{tab:feature_across_tasks}
  \centering
  \begin{tabular}{ l c c c c c c c }
    \toprule
     \textbf{Tasks}   &  \textbf{MFCC}  & \textbf{Spect. Harmonicity} & \textbf{Log HNR} & \textbf{Shimmer} &  \textbf{Spectral Energy} & \textbf{Pitch} & \textbf{Disc. Markers}   \\
     %\hline
                        % & ASD & TD  & ASD & TD  & ASD & TD  & ASD & TD  & ASD & TD  & ASD & TD  & ASD & TD \\
    \hline
    Description~~~   &    0.2409    &   -  &  -   &  -  &   - &    & -     \\
    %Conversation~~~   &    -      &  -   &   -  &  -  &  -  &  -  &  -        \\
    Emotion~~~        &    -0.2504      &  0.2633   &   0.2641  & 0.2439   &   0.2993 &  0.2193  &  -       \\
    Social~~~         &   -   & -    & -    &  -  & -   & -   & -0.2401      \\
    Friends~~~        &  -0.2167        & -   &  -   &  -  &  -  &   - & -         \\
    Loneliness~~~     &  -0.5402     &   -  & -    &  -   &   - &  -  &  -       \\
    %Construction~~~   &   0.2223       &  -   &  0.1815   &  0.1915  & -   &  -  &   -     \\
    %Make-believe~~~   &   0.2973   &     -     &   -  &  -  & -   & -   &  0.2164          \\
    Interactive~~~    & 0.3424  &  -    &  -   & -   & -   &  -0.2601  &  -          \\
    %Demonstration~~~  &  0.2855   & 0.2131 & -   &  -  & 0.2335 & - &   0.2178   \\
    Telling~~~        &  0.2242  &  -   &  -    &  -  & -   & -   &  -     \\
    %Cartoons          &     -     & 0.2014   &    0.2116   &  0.1965  & 0.1930   &  0.1919  &  -      \\
    %Imaginative        &   0.1786      &  -   &   -   &  -  & -   & -   &  -       \\
    %Break              &    0.1997     &  -   &   -   &  -  & -   & -   &  0.1545       \\
    \bottomrule
  \end{tabular}
  
\end{table*}

\section{Method}

We trained  models to analyze the contribution of speech and language features in distinguishing between children with and without ASD during different conversational tasks\footnote{These experiments and the original data collection were reviewed and approved by the Institutional Review Board at the participating institutions.}. 

\subsection{Features}
We used openSMILE~\cite{10.1145/1873951.1874246} to extract acoustic-prosodic features from each turn. In this work, we used the ComPARE 2016 feature set~\cite{schuller2016interspeech}, which extracts 88 low level features and 6373 features derived by applying statistical functions on the low-level features. This tool and a similar feature set have been used before in research on autism diagnosis~\cite{kim2021analyzing}. The low-level features include spectral (Mel Frequency Cepstral Coefficients (MFCC), zero crossing rate), voice quality (local shimmer, jitter, harmonic noise ratio), and prosodic features (loudness, pitch). We z-normalized all these features per participant as the mean age of diagnosed children coincided with the age of changing voice.

We used two sets of language features: Linguistic Inquiry and Word Count (LIWC) features and lexical features motivated by prior research. As prior work has shown correlations between the psychologist's language use and ASD severity~\cite{kumar16b_interspeech}, we calculated these features for each speaker (child and psychologist) for each of the 14 subtasks. 
%We hypothesized that conversation context would influence acoustic-prosodic and language usage patterns in conversation. 
%Since the ADOS-2 consists of subtasks to assess social interaction, the tasks are designed for a variety of contexts including communication, narration, and play.%
To understand the influence of task type and context in predicting the diagnostic group, we also used task as a feature.

LIWC~\cite{pennebaker1999linguistic} has been widely used for predicting outcomes including psychological~\cite{rathner18_interspeech} and cognitive~\cite{asgari2017predicting} functioning and personality~\cite{mairesse2006automatic}. We used a total of 119 LIWC features, including percentage of different parts of speech (e.g. pronouns, articles), punctuation categories (e.g. commas, periods), and psychometric measures (e.g. affect, social, politeness). 

%words:https://github.com/words/hedges
%discourse: https://github.com/sileod/Discovery

Separately, we used a set of 12 lexical features inspired by their use in prior autism research~\cite{song2021natural}. Using the spaCy NLP library~\cite{spaCy} and word dictionaries~\cite{words}, we calculated the relative frequencies of certain parts of speech as well as the number of syllables, hedge words, weasel words, filler words and discourse markers per turn. This resulted in a total of 131 language features. Prior research has shown the importance of appropriate usage of discourse markers (e.g. \textit{and, but, anyway}), hedge words (e.g. \textit{often, usually}), weasel words (e.g. \textit{may, like, possibly}), and filler words (e.g. \textit{so, ok}) for measuring reciprocity and concreteness in conversations for autism research~\cite{yang2021predicting}. 

\subsection{Feature Selection and Classification}
First, we performed correlational analyses to estimate the association between our features and ASD diagnosis outcome. As our diagnosis outcome is binary, we performed point-biserial correlations~\cite{brown2001point}. \rev{To control for any false discovery rate arising from  multiple tests, we applied Holm’s sequential Bonferroni procedure~\cite{10.1111/j.2517-6161.1995.tb02031.x}.}  % We used the features with a statistically significant ($p<0.05$) correlation with diagnosis outcome for each task in our ASD diagnosis model.

Second, to understand the role of different features in distinguishing between children diagnosed with ASD and TD, we used a feature fusion strategy  where we trained a random forest classifier using different combinations of features from different modalities (speech, text, task) extracted from each speaker in the conversation. To identify a minimal set of characteristic features, we used our correlation analysis as a feature selection strategy where we selected features found to be statistically significant ($p<0.05$) for each task into the feature subset. % This subset is used for the classification model taking into account significance level and their order of correlation. 

To ensure generalization for out-of-sample testing, we performed cross-validation by using a leave-$n$-user-out method\footnote{We chose this over 10-fold cross validation, as this ensures our training set does not include information from a child who is also present in our test set.}. We report results averaged over 10 runs, where 80\% (23) of children were randomly selected for training and 20\% (6) for testing in each run. All classification results are reported as accuracy, precision, recall, F1-score and  AUC score. All experiments and models are implemented using the scikit-learn and sciPy libraries with default parameter settings.

\section{Analysis and Results}

%In this section, we report the findings from our experiments to inform our research questions.

\subsection{Correlation Analysis}

Based on the correlation analyses, we found a subset of 28 (out of 88) acoustic-prosodic features, 3 (out of 131) lexical features and 21 (out of 219 total) combination features, extracted from the child conversational participants' turns, to have statistically significant correlations with an ASD diagnosis outcome. In the case of the experimenter, we have 19, 7, and 17 features respectively that have statistically significant correlations with an ASD diagnosis outcome. We present in Table~\ref{tab:feature_across_tasks} the 7 features we find significant most frequently across different tasks. %Among the top 5 features were MFCC [10-14], which is an acoustic measure, which has a positive correlation with ASD for 10 of the subtasks. For tasks of an interview nature (e.g. Conversation, Emotion, Loneliness), it was not a significant indicator, whereas tasks that involved activity by the child (e.eg. Make-believe Play, Construction, Telling). Harmonicity (log HNR) was positively correlated  with diagnosis when talking about loneliness and shimmer when talking about emotion and Loneliness. Spectral energy, a measure of loudness, is also positively correlated when talking about Emotion, Demonstrating a task and act as a cartoon. Pitch (measured by fundamental frequency $F_0$ is positively correlated for Emotion, Loneliness and act as a cartoon. This conforms to the findings in~\cite{bonneh2011abnormal} about role of spectral content and pitch in Autism, but expands under different task context (e.g. negatively correlated when playing with psychologist in Interactive Play).%

In our correlation analysis, we also explored the top features based on significant correlation and strength in each task. To find shared features for both child and psychologist, we compared the top features for the two speakers in the conversational context, which we present in Table~\ref{tab:classification_child_exp_best}. %MFCC is the top feature for tasks involving activity, with a very strong negative association with the Loneliness task (corr. = $-0.5402$). On the other hand, MFCC[13] had a strong positive correlation with the Make-believe task (corr. = $+0.4467$). For child, Speech loudness (Spectral energy) is positively correlated for ASD children when talking about emotion and demonstrating a task. For language features, discourse markers and words related to `Social' category (e.g. \textit{friends, family}) are strong indicators, which are less used in ASD child speech on topics related to social difficulty and unstructured conversation on a topic.%

\subsection{Observations} \textit{Child.} Here, we include some additional observations regarding informative features for different task contexts. When using features from the child conversational participant, in conversations about social difficulties children with ASD used fewer discourse markers ($r$ = $-0.2401$) and fewer words. Children with ASD used fewer pronouns ($r$ = $-0.3703$) and discourse markers ($r$ = $-0.3465$), and their turns were shorter ($r$ = $-0.3299$).
For the make-believe play task, children with ASD used more hedge words ($r$ = $+0.2150$). %For this task as well, apart from MFCC, significant features are language features%
Interestingly, when talking about friends, children with ASD used fewer words from the \textit{tech} category than TD children ($r$ = $-0.1093$).

When talking about emotion, describing a picture, and during the break, the count of words, words per sentence, nouns, syllables and turn duration were all negatively correlated with an ASD diagnosis outcome, which suggest shorter responses from ASD children.
Acoustic-prosodic features such as loudness and pitch were positively correlated with an ASD diagnosis outcome for the cartoons subtask, but negatively correlated for silent play during break.

\textit{Psychologist.} When talking about emotions or friends, the psychologist used fewer discourse markers ($r$ = $-0.2091$), weasel words, words, 
 and syllables, and produced shorter turns, when talking with children with ASD. When talking about loneliness, use of the word \textit{you} ($r$ = $0.3173$) is increased and \textit{i} decreased. The counts of words, words per sentence and pronoun usage (\textit{he/she, we, they}) were all negatively correlated with an ASD diagnosis outcome in conversation, break and emotion subtasks. Overall, language features were found to be more informative in the psychologist data for predicting the diagnosis outcome, whereas only MFCC features among all acoustic-prosodic features were found to be significant. This observation is reflected in our classification results discussed next.

\begin{table*}[thp]
  \caption{Most predictive features for different subtasks with task descriptions. We add the indicator's correlation coefficient with an ASD diagnosis outcome for features extracted from the experimenter's turns when found significant. (C = Child, P = Psychologist)}
  \label{tab:classification_child_exp_best}
  \centering
  \begin{tabular}{ l l l l}
    \toprule
    \textbf{Subtask Name} & \textbf{Task Description} & \textbf{Feature} & \textbf{Corr. Coefficient (C $\vert$ P)}\\
    \midrule
    Description         &   Description of a picture     & MFCC [13] & $+0.2409$ $\vert$ $-0.2064$ ~~~~~~\\
    Conversation        &   Conversation about topics of interest     &  Duration &     ~~~~~~~~~~~~~~~  $\vert$ $-0.2675$     ~~~\\
    Emotions            &   Interview about their feelings      & Spectral energy  & $+0.2993$ ~~~~~~~\\
    Social Difficulties &   Interview about their social lives      & Discourse Markers  & $-0.2401$ ~~~~~\\
    Friends             &  Interview about their relationships      & MFCC [5]    & $-0.2167$ %$\vert$ $-0.19052$% 
    ~~~~~~~~\\
    Loneliness          &   Interview about loneliness     & MFCC [7] & $-0.5402$ ~~~~~~~~~~\\
    Construction        &   Work on a puzzle   & MFCC [14] & $+0.2223$ $\vert$ $-0.2671$ ~~~      \\
    Make-believe Play   &   Child plays with toys alone     & MFCC [13] & $+0.2973$ $\vert$ $+0.4467$ ~~~~~~~~~\\
    Interactive Play    &  As previous (with psychologist)      & MFCC [6] & $+0.3424$ $\vert$ $+0.2603$~~~~~~~~~~\\
    Demonstration       &  Demonstrates how to do a task    & Spectral energy & $+0.2335$ ~~~~~~~\\
    Telling             &  Makes up story based on picture book      & MFCC [13] & $ +0.2242$ ~~~~~\\
    Cartoons            &   Act out a cartoon story     & Duration & $+0.3044$   ~~~~\\
    Imaginative Story   &  Creating a story with small prop items      & MFCC [1] & $+0.1786$ $\vert$ $+0.1765$~~~~~~~~~~~\\
    Break               &   Silent play and unstructured conversation     & MFCC[6] & $+0.1997$ $\vert$ $+0.2014$~~~\\ 
    \bottomrule
  \end{tabular}
  
\end{table*}

%%%%%%%%results before Bonferroni correction, in initial submission%%%%%%%%%
\iffalse
% combined table of before and after feature selection
\begin{table}[th]
  \caption{ASD diagnosis classification from child turns using a random forest classifier trained with different combinations of full and selected features. The number of features used in each model is indicated by N. (A: Acoustic-prosodic, L: Lexical, T: Task category)}
  \label{tab:classification_child}
  \centering
  \begin{tabular}{ l l l l l l l }
    \toprule
    \textbf{Features} & 
                                         \textbf{Acc.} & \textbf{Prec.} & \textbf{Rec.} &
                                         \textbf{F1} &
                                         \textbf{AUC}\\
    \midrule
    A (N = 88)                   & 0.48 & 0.67 & 0.48 & 0.50 & 0.55~~~\\
    A  (N = 57)                  & 0.50 & 0.60 & 0.50 & 0.51 & 0.52~~~\\
    \hline
    L  (N = 131)                    & 0.60 &  0.72 &  0.60 & 0.57 & 0.63~~~             \\
    L  (N = 67)                    &  0.62 & 0.81 & 0.62 & 0.59 & 0.67 ~~~             \\
    \hline
    A + L  (N =219)                    & 0.66 & 0.73 & 0.66 & 0.61 & 0.62~~~      \\
    A + L  (N = 114)                    & 0.64 & 0.77 & 0.64 & 0.59 & 0.63 ~~~      \\
    \hline
    A+L+T (N = 209)      & \textbf{0.74} & 0.85 & \textbf{0.74} &  \textbf{0.72} & 0.76 ~~~\\
    A+L+T  (N = 115)    & 0.72 & \textbf{0.87} & 0.72 & 0.71 & \textbf{0.79}       ~~~\\
    \bottomrule
  \end{tabular}
  
\end{table}
\fi

%%%%%% Corrected results after Bonferroni correction %%%%%%
\begin{table}[th]
  \caption{ASD diagnosis classification from child turns using a random forest classifier trained with different combinations of full and selected features. The number of features used in each model is indicated by N. Best results are boldfaced. (A: Acoustic-prosodic, L: Lexical, T: Task category)}
  \label{tab:classification_child}
  \centering
  \begin{tabular}{ l l l l l l l }
    \toprule
    \textbf{Features} & 
                                         \textbf{Acc.} & \textbf{Prec.} & \textbf{Rec.} &
                                         \textbf{F1} &
                                         \textbf{AUC}\\
    \midrule
    A (N = 88)                   & 0.48 & 0.67 & 0.48 & 0.50 & 0.55~~~\\
    A  (N = 28)                  & 0.51 & 0.65 & 0.51 & 0.55 & 0.56~~~\\
    \hline
    L  (N = 131)                    & 0.60 &  0.72 &  0.60 & 0.57 & 0.63~~~             \\
    L  (N = 3)                    &  0.61 & 0.76 & 0.61 & 0.55 & 0.62 ~~~             \\
    \hline
    A + L  (N =219)                    & 0.66 & 0.73 & 0.66 & 0.61 & 0.62~~~      \\
    A + L  (N = 21)                    & 0.62 & 0.81 & 0.62 & 0.58 & 0.66 ~~~      \\
    \hline
    A+L+T (N = 220)      & 0.74 & \textbf{0.85} & 0.74 &  0.72 & \textbf{0.76} ~~~\\
    A+L+T  (N = 22)    & \textbf{0.76} &  0.84 & \textbf{0.76} & \textbf{0.75} & {0.71}      ~~~\\
    \bottomrule
  \end{tabular}
  
\end{table}

\subsection{Classification}

We use the findings of our correlation analyses to inform our feature selection strategy when training classifiers to distinguish between children with ASD and TD children. We first trained a classifier with all features from each speaker turn, and then separately trained classifiers with selected features from different combinations of feature sets. We used random forests because their results are highly interpretable by psychologists. Our goal was to find a minimal feature set for the purpose of aiding diagnosis. % While other classification models may improve the classifier's ability to distinguish ASD and TD children, it is out of the scope of our goal in this paper.

Our classification results for child turn data are presented in Table~\ref{tab:classification_child}. For each feature type, we trained with full and selected feature sets (the number of features, $N$, is indicated in parentheses). We found language features were better indicators for the diagnosis outcome than acoustic-prosodic features. However, the combination of features from the two modalities gave improved accuracy ($F1=0.61$). \rev{When we trained with features selected using point-biserial correlation analysis, the number of features used in the model was reduced by nearly 80\% for the A, A+L and A+L+T settings, while classification accuracy improved or remained unchanged.} After including the subtask category as a categorical feature, we obtained the best classification results with both full and selected feature sets. % This can be explained by the notion that task category as a feature helps the model to differentiate between the two diagnostic groups in different subtask contexts.

We observed similar findings when using the psychologist's turn data which we present in Table~\ref{tab:classification_exp}. \rev{Using only acoustic-prosodic features (A) and only text (T), the model trained on the psychologist's turns provided results similar to the one trained on the childs' turns. We observe that the classification results remained unchanged or improved for all settings (A, L, A+L, A+L+T) when training with nearly 80\% fewer features selected through the correlation analysis feature selection procedure, an observation similar to the results from child's turns.}

%However, for language features ($L$), this observation did not hold. % One explanation for this can be that adult conversation context can span over multiple turns instead of a single turn, which we plan to explore in future works. From correlational analyses, we selected a slightly higher number of features, and performance remained unchanged or improved with this minimal set. When including Task Category ($L$) as a feature, we obtained the best performance when selected features included 36\% fewer features ($A + L+ T$).

We also applied dimensionality reduction (PCA) as an alternate feature selection strategy and compared this with the correlation-based selection strategy. The number of principal components was chosen to preserve 95\% variance in the data. We found that this did not improve performance compared to using the full feature set. One reason could be that we fit a PCA model to the training data and transformed the testing data using the fitted model. Since our training and testing data come from different children, applying PCA did not help in this case.

\textbf{Ethical discussion and limitations.} Our goal is to provide AI decision tools to help assessors more effectively use ADOS for ASD diagnosis. It would be unethical to deploy the machine learning approach outlined here on its own (without human judgment). Also, it would be unethical to deploy such an approach in a non-clinical setting. In order to ensure no misuse of our findings, we do not plan to release the trained models or raw data to the public. %We also do not plan to release raw data, only calculated features. We will update the final submission to reflect this.%

For our experiments, we used entirely automatic preprocessing. No stage of preprocessing is perfect. In particular, there are many cases where pyannote missed speaker changes. %Also, Whisper is less accurate on child speech than it is on adult speech.%
Our experimental results could be improved with manual diarization and transcription, but automatic diarization and transcription are more practical in our target setting.

Another limitation of this work is that our data includes only two diagnostic groups (ASD and TD) and the language of the conversations is English. Whether the informative features for ASD are also applicable for other co-occurring neuro-developmental disorders (e.g. ADHD) and other languages is a direction for future work. Finally, the current data set only included one psychologist, so we do not know how our findings would generalize to other practitioners.

%%%%%%%%results before Bonferroni correction, in initial submission%%%%%%%%%
\iffalse
\begin{table}[th]
  \caption{ASD diagnosis classification from the assessor's turns using a random forest classifier trained with different combinations of full and selected features. The number of features used in each model is indicated by N. (A: Acoustic-prosodic, L: Lexical, T: Task category)}
  \label{tab:classification_exp}
  \centering
  \begin{tabular}{ l l l l l l l }
    \toprule
    \textbf{Features} & \textbf{Acc.} & \textbf{Prec.} & \textbf{Rec.} &
                                         \textbf{F1} &
                                         \textbf{AUC}\\
    \midrule
    A  (N = 88)                  & 0.49 & 0.60 & 0.49 & 0.50 & 0.52~~~\\
    A  (N = 53)                  &  0.51 & 0.64 & 0.51 & 0.52 & 0.57 ~~~\\
    \hline
    L  (N = 131)                    & 0.56 & 0.79 & 0.56 & 0.54 & 0.63 ~~~             \\
    L  (N = 89)                    &  0.52 & \textbf{0.80} & 0.52 & 0.45 & 0.61  ~~~             \\
    \hline
    A + L  (N = 219)        & 0.55 & 0.70 & 0.55 & 0.47 & 0.59~~~      \\
    A + L  (N = 141)        & 0.57 & 0.71 & 0.57 & 0.49 & 0.58 ~~~      \\
    \hline
    A+L+T  (N = 220)      &  0.59 & \textbf{0.80} & 0.59 & 0.56 & 0.64 ~~~\\
    A+L+T   (N = 142 )                     &  \textbf{0.66} & 0.72 & \textbf{0.66} & \textbf{0.60} & \textbf{0.67}      ~~~\\
    
    \bottomrule
  \end{tabular}
  
\end{table}
\fi

%%%%%%%%corrected results after Bonferroni correction%%%%%%%%%
\begin{table}[th]
  \caption{ASD diagnosis classification from psychologist's turns using a random forest classifier trained with different combinations of full and selected features. The number of features used in each model is indicated by N. Best results are boldfaced. (A: Acoustic-prosodic, L: Lexical, T: Task category)}
  \label{tab:classification_exp}
  \centering
  \begin{tabular}{ l l l l l l l }
    \toprule
    \textbf{Features} & \textbf{Acc.} & \textbf{Prec.} & \textbf{Rec.} &
                                         \textbf{F1} &
                                         \textbf{AUC}\\
    \midrule
    A  (N = 88)                  & 0.49 & 0.60 & 0.49 & 0.50 & 0.52~~~\\
    A  (N = 19)                  &  0.51 & 0.63 & 0.51 & 0.51 & 0.56 ~~~\\
    \hline
    L  (N = 131)                    & 0.56 & 0.79 & 0.56 & 0.54 & 0.63 ~~~             \\
    L  (N = 7)                    &  0.63 & 0.82 & 0.63 & 0.60 & 0.66  ~~~             \\
    \hline
    A + L  (N = 219)        & 0.55 & 0.70 & 0.55 & 0.47 & 0.59~~~      \\
    A + L  (N = 17)        & 0.61 & 0.74 & 0.61 & 0.53 & 0.56 ~~~      \\
    \hline
    A+L+T  (N = 220)      &  0.59 & 0.80 & 0.59 & 0.56 & 0.64 ~~~\\
    A+L+T   (N = 18 )                     &  \textbf{0.68} & \textbf{0.83} & \textbf{0.68} & \textbf{0.65} & \textbf{0.67}      ~~~\\
    
    \bottomrule
  \end{tabular}
  
\end{table}

%\section{Discussion}

\section{Conclusions and Future Work}

Early diagnosis and interventions are critical to positive outcomes for autism. There is a growing need for identifying behavioral indicators that can help clinicians to diagnose ASD. Speech and language features have been extensively studied in prior works to distinguish between children with ASD and TD children. In this work, we explored the diagnostic informativeness of different acoustic-prosodic and language features across multiple diagnostic conversational contexts, separately examining data from child and psychologist turns. We used acoustic-prosodic indicators %(voice quality, loudness, etc.)% 
and language indicators %(linguistic word counts from LIWC)%
to identify the most informative features for different conversational tasks administered during the ADOS-2. Our results from correlational and classification experiments suggest the presence of informative indicators that are shared across different contexts and co-appear in conversational turns from both speakers. 

We are currently working to confirm these results via manually corrected speaker diarization and transcription. In future work, we would like to extend this analysis to multimodal conversational data. We also plan to examine modeling approaches that take into account more of the conversational context, for example, whether the child conversation participant constructs contextually appropriate dialog acts, or the ways in which child and assessor align in acoustic-prosodic or linguistic ways during conversation.

\section{Acknowledgements}
We would like to thank and acknowledge all our participants and their families. The data set was obtained thanks to support by the National Institute of Mental Health and the National Institutes of Health: awards R21MH094659 and F31MH108331. %(redacted). %R21MH094659 and F31MH108331%.

\newpage 

\bibliographystyle{IEEEtran}
\bibliography{references}

\end{document}